\documentclass[10pt,twocolumn,letterpaper]{article}

\newif\ifdraft
\drafttrue

\usepackage[margin=0.75in,columnsep=0.25in]{geometry}
\usepackage{mathptmx}
\usepackage[T1]{fontenc}
\usepackage{amsmath,amssymb}
\usepackage{graphicx}
\usepackage{booktabs}
\usepackage{xcolor}
\usepackage[font=small,labelfont=bf]{caption}
\usepackage{microtype}
\usepackage{cite}
\usepackage[colorlinks=true,linkcolor=black,citecolor=blue!50!black,urlcolor=blue!50!black]{hyperref}

\ifdraft
  
  \newcommand{\tbd}[1]{\textcolor{orange!85!black}{\textbf{[TBD: #1]}}}
\else
  
  \newcommand{\tbd}[1]{}
\fi

\title{\Large\bfseries Smoothing and Learning for Movement-Robust\\
RSS-Based Spoofing Detection}

\author{
\normalsize Usama Mir (Senior Member IEEE)\\
\normalsize Wilfrid Laurier University, Ontario, Canada\\
\normalsize umir@wlu.ca
}
\date{}

\begin{document}
\maketitle
\begin{abstract}
Physical-layer spoofing detection identifies an attacker who impersonates a
legitimate wireless user by checking whether transmissions claiming the same
identity originate from different locations, using the received signal
strength (RSS) measured at several receivers. A recent scheme by Romero et al.\
makes this approach robust to user movement by combining a neural-network
position-change detector (PCD) with community detection on a graph of frames
and a fixed threshold on a hand-crafted test statistic. I study two
modifications to this scheme. First, I smooth the sequence of PCD scores
with a median or Gaussian temporal filter before thresholding. Second, I
replace the fixed decision threshold with a supervised classifier operating
on six statistical features of the region sequence produced by community
detection. On real indoor RSS measurements from 52 transmitter locations,
Gaussian filtering raises PCD accuracy from 0.870 to 0.921 when the PCD is
trained on 10 locations and from 0.905 to 0.925 when trained on 40.
On synthetic attack and no-attack sequences built from the same
measurements, evaluated on a held-out test set of 1{,}679 sequences, the best
classifier (XGBoost) reaches an accuracy of 0.949
and an $F_1$ score of 0.940, and random forests reach an AUC of 0.99. The
detector separates static attackers almost perfectly but misses about a third
of attacks in which two transmitters move simultaneously, which produce region
sequences similar to those of a single moving user.
\end{abstract}

\noindent\textbf{Index Terms}---Spoofing detection, physical layer security,
received signal strength, user mobility, community detection, machine learning.

\section{Introduction}

Wireless networks, from cellular systems such as LTE~\cite{mir2013lte} to
cognitive radio networks in which users share spectrum
dynamically~\cite{mir2011ctmc,mir2018handoff}, are open to impersonation: an
attacker who obtains a legitimate user's credentials, or forges identifiers such as MAC addresses,
can transmit on that user's behalf~\cite{kolias2016,vanhoef2016,tippenhauer2011}.
Cryptographic authentication remains essential, but it cannot detect an
attacker who holds valid credentials. Physical layer security (PLS) offers a
complementary check based on properties of the radio channel that are
difficult to forge~\cite{marabissi2022,illi2024}. Received signal strength
(RSS)~\cite{yang2013} and channel state information
(CSI)~\cite{martins2023,li2021} depend on the transmitter's location, so
transmissions that claim the same identity but arrive from different locations
indicate a spoofing attack.

The difficulty is that a legitimate user who moves also produces changing RSS
signatures. Romero, Ha and Gerstoft~\cite{romero2024} addressed this by
detecting whether frames come from \emph{concurrent} distinct locations rather
than from \emph{any} change in location. Their scheme uses a deep neural
network as a position-change detector (PCD) for pairs of frames, builds a
graph of frames from the PCD decisions, partitions it with community
detection, and declares an attack when a test statistic computed from the
resulting sequence of regions exceeds a fixed threshold.

Two parts of this pipeline are sensitive to noise. Individual PCD decisions
are made independently for each pair of frames, so fluctuations in RSS caused
by multipath produce isolated errors. The final decision is a fixed threshold
on a single scalar statistic, which may not capture the variety of region
patterns produced by different movement behaviors.

In this paper I study two modifications that address these points:
\begin{itemize}
  \item a \textbf{temporal filter} (median or Gaussian) applied to the
        sequence of PCD scores before thresholding, which exploits the fact
        that consecutive frames usually come from the same transmitter; and
  \item a \textbf{learned decision rule}: six statistical features of the
        region sequence, including a normalized version of the original test
        statistic, are fed to a supervised classifier. I compare six
        classifiers.
\end{itemize}
I evaluate both on the real indoor RSS dataset of~\cite{romero2024}, using
synthetic sequences that emulate static and moving users with and without an
attacker. I also report where the approach fails: attacks by two moving
transmitters remain hard to detect.

\section{Related Work}

Illi et al.~\cite{illi2024} survey PLS for authentication, confidentiality
and malicious-node detection in IoT networks, and identify RSS and CSI as key
features for lightweight authentication. Yang et al.~\cite{yang2013} and
Gajbhiye and Daruwala~\cite{gajbhiye2016} use RSS to detect spoofing in
IEEE~802.11 networks and to estimate the number of attackers by clustering.
Pinto et al.~\cite{pinto2018} combine a moving-average filter, $k$-means
clustering and a $k$-NN classifier to detect spoofing in wireless sensor
networks, including the hard case in which the attacker and the legitimate
node are equidistant from a landmark. Marabissi et al.~\cite{marabissi2022}
combine PLS features with machine learning for IoT node authentication.
CSI-based detectors have been proposed for MIMO systems~\cite{martins2023}
and for dynamic channels with imperfect CSI~\cite{li2021}.

Most of these methods assume static users, so a legitimate user's movement
is indistinguishable from an attack. Romero et al.~\cite{romero2024} proposed
the movement-robust scheme on which I build, and Ha and
Romero~\cite{ha2024} later replaced the community-detection stage with a
graph neural network. My work stays with the lightweight community-detection
pipeline and asks whether simple temporal smoothing and conventional
classifiers improve it.

\section{Background: Movement-Robust Spoofing Detection}

\subsection{Problem Formulation}
A set of transmitters sends frames to $M$ receivers, each of which measures
RSS. A frame is the vector $f[j] = (\mathrm{RSS}_{1},\dots,\mathrm{RSS}_{M})$
of measurements for the $j$th transmission. Given a sequence
$F = \{f[0],\dots,f[J-1]\}$ of frames that claim the same identity, let
$F_L$ be the frames sent by the legitimate user and $F_A = F\setminus F_L$
those sent by an attacker. The system is safe if $F_A=\emptyset$ and under
attack otherwise. The legitimate user and the attacker may each be static or
moving.

\subsection{Baseline Scheme~\cite{romero2024}}
\textbf{Position-change detector.} A DNN takes a pair of frames
$(f, f')$ and outputs a score; thresholding the score gives
$d^{\mathrm{PCD}}(f,f') \in \{\mathrm{True},\mathrm{False}\}$, where True means
the frames come from the same location.

\textbf{Spoofing detector.} A graph $g$ is built with one node per frame
and an edge between $f[j]$ and $f[j']$ whenever
$d^{\mathrm{PCD}}(f[j],f[j'])=\mathrm{True}$. Community detection on $g$
assigns each frame to a region, giving a region sequence
$r[0],\dots,r[J-1]$. A legitimate moving user tends to move through regions
progressively, whereas concurrent transmissions from two locations make the
sequence return repeatedly to earlier regions. The test statistic rewards such
returns, weighting recent returns more heavily:
\begin{gather*}
  T^{\mathrm{SD}} = \sum_{j=3}^{J-1} w[j],\qquad
  w[j] = \begin{cases} 1/\tilde w[j] & \tilde w[j] \neq 0\\ 0 & \text{otherwise}\end{cases}\\
  \tilde w[j] = \bigl|\{r[j^*+1],\dots,r[j-1]\}\bigr|,
\end{gather*}
where $j^*$ is the largest index $j^*<j$ with $r[j^*]=r[j]$ and $|\cdot|$
counts distinct regions. An attack is declared when
$T^{\mathrm{SD}}>\gamma^{\mathrm{SD}}$ for a fixed threshold
$\gamma^{\mathrm{SD}}$.

\textbf{Sensitivity to noise.} Both stages depend on thresholds. Isolated PCD
errors caused by RSS fluctuations add spurious edges or remove true ones,
which changes the community structure, and the single fixed threshold
$\gamma^{\mathrm{SD}}$ must cover all movement patterns.

\section{Proposed Modifications}

\begin{figure}[t]
  \centering
  \includegraphics[width=\columnwidth]{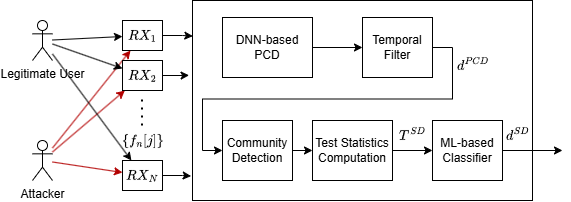}
  \caption{Modified detection pipeline. A temporal filter smooths the PCD
  scores, and a machine-learning classifier replaces the fixed threshold on
  the test statistic.}
  \label{fig:pipeline}
\end{figure}

Fig.~\ref{fig:pipeline} shows the modified pipeline. The DNN-based PCD and
community detection are unchanged; a temporal filter is inserted after the
PCD, and the threshold on $T^{\mathrm{SD}}$ is replaced by a classifier.

\subsection{Temporal Filtering of PCD Scores}
In practice, a legitimate user and an attacker do not alternate frame by
frame; each transmits bursts of tens or hundreds of consecutive frames. The
PCD score of one pair is therefore strongly predictive of the scores of
neighboring pairs, and an isolated score that disagrees with its neighbors is
likely an error. Let $x_1,\dots,x_n$ be the time-ordered sequence of PCD
scores. I consider two filters before thresholding.

\textbf{Median filter} of window size $q$:
\begin{equation}
  \tilde x_i = \operatorname{median}\left(x_{i-\lfloor q/2\rfloor},\dots,x_{i+\lfloor q/2\rfloor}\right).
\end{equation}

\textbf{Gaussian filter}~\cite{kopparapu2011}, which weights nearby scores
more than distant ones:
\begin{equation}
  \tilde x_i = \sum_{k=-r}^{r} G(k)\,x_{i+k},\quad
  G(k) \propto \exp\!\left(-\frac{k^2}{2\sigma^2}\right),\quad r = 3\sigma,
\end{equation}
with $G$ normalized to sum to one. Filters are applied only at test time; the
DNN is trained on unfiltered pairs.

\subsection{Learned Spoofing Decision}\label{sec:learned}
\begin{figure}[t]
  \centering
  \includegraphics[width=0.62\columnwidth]{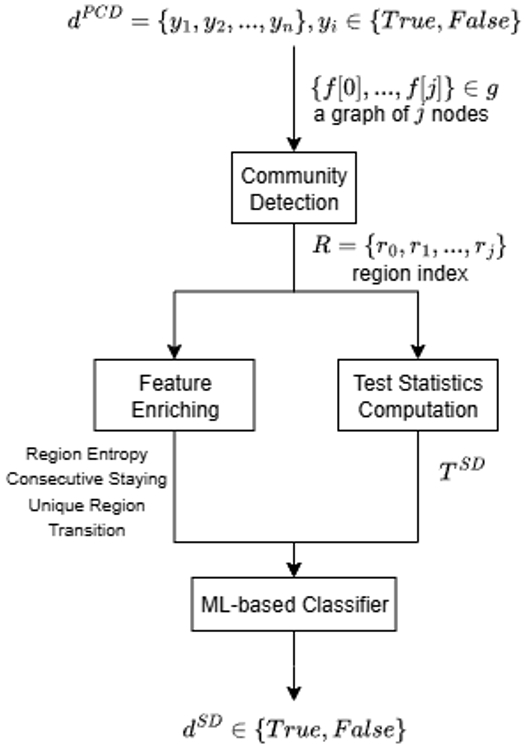}
  \caption{Learned spoofing decision: features of the region sequence are
  combined with the test statistic and passed to a classifier.}
  \label{fig:sdflow}
\end{figure}

Instead of thresholding $T^{\mathrm{SD}}$ alone, I describe each region
sequence $r[0],\dots,r[J-1]$ with six features:
\begin{itemize}
  \item \textbf{TSD ratio}: $T^{\mathrm{SD}}$ divided by the sequence length;
  \item \textbf{transition ratio}: number of region changes
        ($r[j]\neq r[j-1]$) divided by the sequence length;
  \item \textbf{unique ratio}: number of distinct regions divided by the
        sequence length;
  \item \textbf{max stay}: the longest run of consecutive frames in one region;
  \item \textbf{mean stay}: the average length of such runs;
  \item \textbf{region entropy}: the entropy of the distribution of frames
        over regions.
\end{itemize}
These features summarize how often the sequence changes region, how many
regions it visits, and how evenly it spreads over them. A classifier trained
on labeled sequences then outputs the attack decision $d^{\mathrm{SD}}$
(Fig.~\ref{fig:sdflow}). I compare logistic regression, a decision tree, a
random forest, a support vector machine (SVM), $k$-nearest neighbors
($k$-NN), and XGBoost~\cite{chen2016}, all trained on the same features and
the same train/test split.

\begin{figure}[t]
  \centering
  \includegraphics[width=0.85\columnwidth]{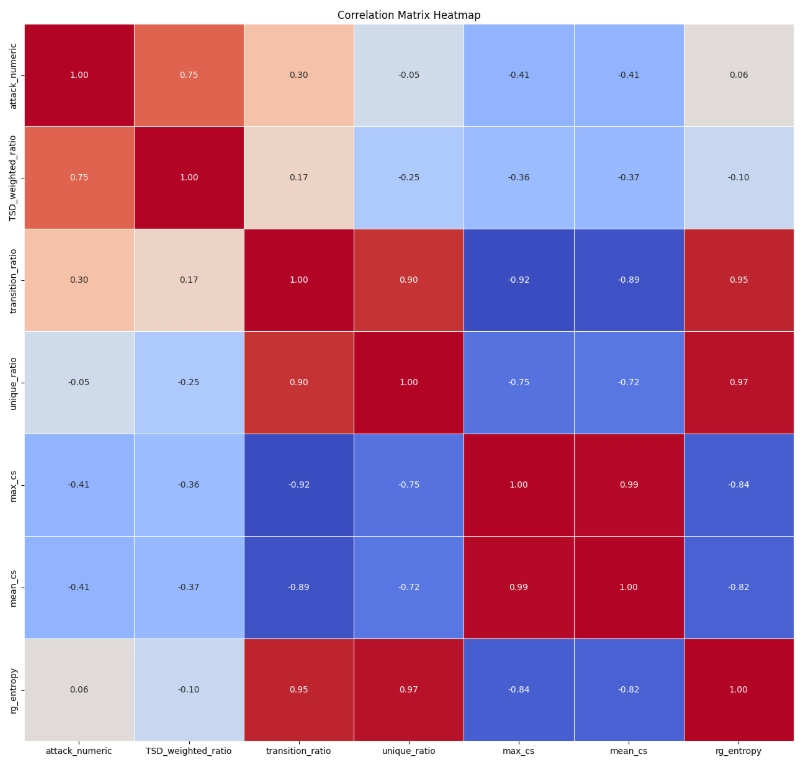}
  \caption{Pearson correlations between the six features and the attack
  label (\texttt{attack\_numeric}).}
  \label{fig:corr}
\end{figure}

Fig.~\ref{fig:corr} shows the correlation of the features with the attack
label. The TSD ratio is by far the most informative single feature
(correlation 0.75), which confirms the value of the original statistic. The
transition ratio is moderately correlated with the label (0.30), and both
stay-length features are negatively correlated with it ($-0.41$): attacks
break the sequence into shorter stays. The unique ratio and region entropy
are almost uncorrelated with the label on their own ($|\rho|<0.1$) and are
strongly correlated with the transition ratio; max and mean stay are nearly
collinear (0.99). The features are therefore largely redundant apart from the
TSD ratio, which helps explain why all classifiers perform similarly
(Section~\ref{sec:results}).

\section{Experimental Setup}

\subsection{Dataset}
I use the RSS dataset of Romero et al.~\cite{romero2024}, collected indoors
on the campus of the University of California, San Diego. A transmitter was
placed at 52 locations along four parallel lines, and four receivers with
four antennas each recorded 4{,}888 samples per location, giving 16 RSS
values per frame.

\subsection{Position-Change Detector}
Following~\cite{romero2024}, the DNN is trained on pairs of frames drawn
from a subset of the 52 locations. Half of the pairs are drawn from the same location and
half from two different locations. To evaluate the temporal filter, I
additionally require same-location pairs to consist of temporally adjacent
frames, $(f_x[j], f_x[j+1])$, so that the scores form a sequence in which
neighboring pairs are close in time. I train on 10 and on 40 locations and
report pair-classification accuracy.

\subsection{Synthetic Spoofing Scenarios}
\begin{figure}[t]
  \centering
  \includegraphics[width=\columnwidth]{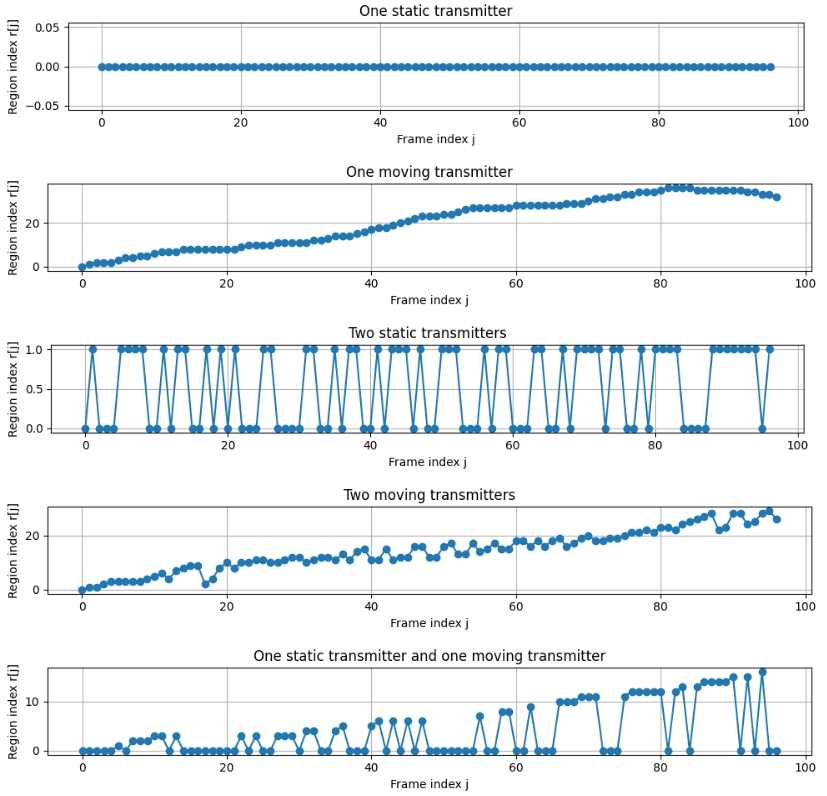}
  \caption{Example region sequences produced by community detection for the
  five scenarios.}
  \label{fig:regions}
\end{figure}

To train and test the spoofing detector I build frame sequences from the
measured data under five scenarios (Fig.~\ref{fig:regions}):
\begin{itemize}
  \item \textbf{1-Sta} (legitimate): all frames from one location $x$.
  \item \textbf{1-Mov} (legitimate): the user starts at location $x_i$; each
        next frame comes from $x_i$ or $x_{i+1}$ with equal probability, so
        the user advances gradually, reversing direction at the last
        location.
  \item \textbf{2-Sta} (attack): each frame comes at random from one of two
        fixed locations $x_1 \neq x_2$.
  \item \textbf{2-Mov} (attack): two transmitters each move as in 1-Mov,
        and each frame comes at random from one of them.
  \item \textbf{Mix} (attack): one static and one moving transmitter.
\end{itemize}
The sequences are processed by the PCD and community detection, and the six
features of Section~\ref{sec:learned} are computed for each resulting region sequence.
The example sequences in Fig.~\ref{fig:regions} contain 97 frames. The
classifiers are trained on part of the sequences and evaluated on a held-out
test set of 1{,}679 sequences: 954 legitimate (314 1-Sta, 640 1-Mov) and 725
attack (343 2-Sta, 78 2-Mov, 304 Mix). Because a moving user is the harder
legitimate case, 1-Mov makes up two thirds of the legitimate test
sequences.

Fig.~\ref{fig:regions} illustrates the patterns the classifier must
separate. A static user stays in one region, and a moving user advances
steadily through regions. Two static transmitters alternate between two
regions, and a static-plus-moving attack keeps returning to the static
transmitter's region. Two moving transmitters, however, produce a sequence
that also trends upward with small backward jumps, which resembles a single
moving user.

\section{Results}\label{sec:results}

\subsection{Effect of Temporal Filtering on the PCD}
Table~\ref{tab:pcd} shows PCD accuracy with and without filtering. Both
filters improve on the unfiltered DNN. The gain is largest when the DNN is
trained on only 10 locations, where Gaussian filtering raises accuracy by
5.1 percentage points; with 40 training locations the unfiltered DNN is
already more accurate and the gain shrinks to 2.0 points. The Gaussian filter
is slightly better than the median filter in both cases. This is useful in
practice because collecting training data at many locations is the main cost
of deploying the scheme in a new environment.

\begin{table}[t]
  \centering
  \caption{PCD accuracy with and without temporal filtering}
  \label{tab:pcd}
  \small
  \begin{tabular}{lcc}
    \toprule
    & \multicolumn{2}{c}{Training locations}\\
    \cmidrule(lr){2-3}
    & 10 & 40 \\
    \midrule
    DNN (no filter)~\cite{romero2024} & 0.870 & 0.905 \\
    DNN + median filter        & 0.905 & 0.922 \\
    DNN + Gaussian filter      & \textbf{0.921} & \textbf{0.925} \\
    \bottomrule
  \end{tabular}
\end{table}

\subsection{Classifier Comparison}
\begin{figure}[t]
  \centering
  \includegraphics[width=0.85\columnwidth]{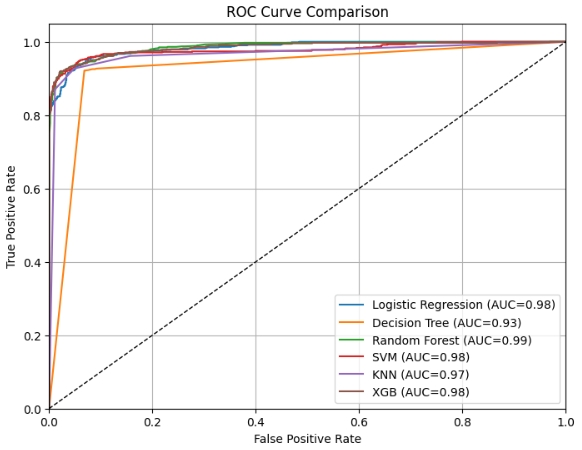}
  \caption{ROC curves of the six classifiers.}
  \label{fig:roc}
\end{figure}

Table~\ref{tab:clf} compares the six classifiers on the test sequences, with
attack as the positive class. All achieve accuracy between 0.927 and 0.949,
consistent with the redundancy of the features noted above. XGBoost has the
highest accuracy (0.949), $F_1$ (0.940) and Matthews correlation coefficient
(0.896); random forests are close behind and have the highest AUC (0.99,
Fig.~\ref{fig:roc}). The SVM has the lowest false-positive rate (1.9\%) but
misses 10\% of attacks, as does logistic regression; the decision tree is the
weakest overall.

\begin{table}[t]
  \centering
  \caption{Spoofing-detector results by classifier (1{,}679 test sequences;
  attack = positive class; AUC read from Fig.~\ref{fig:roc})}
  \label{tab:clf}
  \small
  \setlength{\tabcolsep}{3.5pt}
  \begin{tabular}{lcccccc}
    \toprule
    Classifier & Acc. & Prec. & Rec. & FPR & $F_1$ & AUC \\
    \midrule
    Logistic reg.  & 0.938 & 0.953 & 0.899 & 3.4\% & 0.925 & 0.98 \\
    Decision tree  & 0.927 & 0.911 & 0.921 & 6.8\% & 0.916 & 0.93 \\
    Random forest  & 0.947 & 0.944 & \textbf{0.932} & 4.2\% & 0.938 & \textbf{0.99} \\
    SVM            & 0.946 & \textbf{0.973} & 0.899 & \textbf{1.9\%} & 0.935 & 0.98 \\
    $k$-NN         & 0.941 & 0.936 & 0.927 & 4.8\% & 0.931 & 0.97 \\
    XGBoost        & \textbf{0.949} & 0.953 & 0.927 & 3.5\% & \textbf{0.940} & 0.98 \\
    \bottomrule
  \end{tabular}
\end{table}

\subsection{Performance by Scenario}
Table~\ref{tab:scen} breaks down the XGBoost results by scenario. Static
cases are classified perfectly: no static user is flagged and every
two-static-transmitter attack is detected. A moving legitimate user is
wrongly flagged in 5.2\% of sequences, and 8.6\% of static-plus-moving attacks
are missed. The main weakness is the 2-Mov scenario, in which 34.6\% of attacks
are missed. As Fig.~\ref{fig:regions} suggests, two transmitters that move
gradually produce region sequences with a progressive trend similar to a
single moving user, and the region-level features I use cannot reliably
separate the two.

\begin{table}[t]
  \centering
  \caption{XGBoost results by scenario}
  \label{tab:scen}
  \small
  \begin{tabular}{llrc}
    \toprule
    Scenario & Class & Sequences & Correct \\
    \midrule
    1-Sta & legitimate & 314 & 100.0\% \\
    1-Mov & legitimate & 640 & 94.8\% \\
    2-Sta & attack     & 343 & 100.0\% \\
    2-Mov & attack     & 78  & 65.4\% \\
    Mix   & attack     & 304 & 91.4\% \\
    \bottomrule
  \end{tabular}
\end{table}

\section{Discussion and Limitations}

\textbf{No direct comparison with the fixed threshold.} I compare six
classifiers with each other but have not yet evaluated the original
fixed-threshold detector on the same synthetic sequences. The high
correlation of the TSD ratio with the label suggests that the original
statistic alone is already a strong detector, so the gain from the learned
decision rule remains to be quantified. This comparison is the most
important next step.

\textbf{Synthetic scenarios from one environment.} All sequences are
synthesized from measurements in a single indoor environment. Real movement
is continuous and receivers observe fading that changes over time, which the
random draws between neighboring measurement locations only approximate. I
also assume that consecutive location indices are physically adjacent. I
have not evaluated the method in a second environment, so I make no claim
about cross-environment generalization.

\textbf{Two moving transmitters.} Region-sequence features capture the order
in which regions are visited but not the RSS values themselves. Separating one
mover from two may require features computed directly from the RSS
trajectories or from CSI.

\textbf{Burst assumption.} The temporal filter assumes that each transmitter
sends bursts of consecutive frames. An attacker who interleaves frames finely
with the legitimate user would weaken this assumption, although such
interleaving would also make the attack easier to detect in the region
sequence. The size of the filtering gain also depends on how
strongly consecutive scores are correlated in the evaluation data, so it
should be confirmed on continuous recordings of real transmissions.

\section{Conclusion}

I studied two modifications to a movement-robust RSS-based spoofing
detector. A Gaussian temporal filter on the position-change detector's scores
improves its accuracy from 0.870 to 0.921 when only 10 training locations are
available, reducing the measurement effort needed for deployment. Replacing
the fixed threshold with a classifier on six region-sequence features yields
accuracies of 0.927 to 0.949 across six classifiers, with perfect separation
of static scenarios. Attacks by two moving transmitters remain the main
failure mode, with 34.6\% missed. Future work will compare the learned
decision rule directly with the original threshold, evaluate in other
environments with real movement, and add signal-level or CSI features to
distinguish one moving user from two.

\section*{Code and Data Availability}
The RSS measurements are from~\cite{romero2024}.

\section*{Use of Generative AI}
The manuscript text was revised and edited with the assistance of an AI
language model (Claude, Anthropic). The author reviewed all content and takes
full responsibility for it.


\end{document}